\documentclass[aps,12pt,final,notitlepage,oneside,onecolumn,showpacs,
	nobibnotes,nofootinbib,superscriptaddress,centertags,showkeys]
{revtex4-1}

\usepackage{amsmath}
\usepackage{amsfonts}
\usepackage[dvips]{graphicx}

\usepackage[cp1251]{inputenc}
\usepackage[T2A]{fontenc}

\usepackage{amssymb}

\makeatletter
\renewcommand \thefigure{\@arabic\c@figure}
\renewcommand \thetable{\@arabic\c@table}
\def\place@bibnumber@inl#1{#1.}%
\def\thesection       {\arabic{section}}
\def\p@section        {}
\def\thesubsection    {\thesection.\arabic{subsection}}
\def\p@subsection     {\thesection.}

\def\p@subsubsection  {\thesection\,\thesubsection\,}

\begin{document}

	\title{The geometric interpretation of the cosmological repulsion forces}
	\author{\firstname{A.~V.}~\surname{Klimenko}}
	\email{alklimenko@gmail.com}
	\affiliation{<<Business and Technology>>, Chelyabinsk, Russia}
	
	\author{\firstname{V.~A.}~\surname{Klimenko}}
	\affiliation{<<Chelyabinsk State University>>, Chelyabinsk, Russia}
	\affiliation{<<Business and Technology>>, Chelyabinsk, Russia}
		
\begin{abstract}
It seems likely that the generalized Einstein equations are not complete and only partly account for the effect on the Universe dynamics of that part of the energy of the space environment the change of which is purely geometric. There are offered the generalized Einstein's equations, describing not only the gravity forces, but also the cosmological forces of repulsion, which are geometric in their nature.

The generalized Einstein equations are used to derive the cosmological Friedman's equations describing the dynamics of a homogeneous isotropic universe with the influence of cosmological repulsion forces. We propose a cosmological model of the universe based on these equations. Application of the model for explanation of the observations is considered here.

\bigskip

\end{abstract}	
\maketitle

\newpage
	\tableofcontents
	
\section{Introduction}
\label{vvedenie}
Until rather recently it was believed that the dynamics of the universe is determined by the forces of gravity (see, e.g., \cite{1,2,3,3.1,3.2}). Currently, there are numerous observational data to explain which the assumption on the existence of cosmological repulsion forces is used. It is believed that the dynamics of the homogeneous isotropic universe is largely determined by the influence of these forces (see, e.g., \cite{4,5,6,7,8,9,10}).
	
The explanations are known of the cosmological repulsion forces within the framework of the Einstein equations. Their essence is the following. For the homogeneous isotropic universe the Einstein equations are converted to the cosmological Friedman equations. They can be written as \cite{1}:
\begin{equation}
	\label{da}
		\frac{d \varepsilon}{da}+3 (\varepsilon + P) \frac 1a =0,
	\end{equation}
\begin{equation}
	\label{d2a}
		\ddot{a} = - \frac 43 \pi G \frac{a}{c^2} (\varepsilon + 3P),
	\end{equation}
where $\varepsilon$ and $P$~--- energy density and pressure of the space environment, respectively; $a(t)$~--- radius of the three-dimensional space curvature.

From \eqref{d2a} we see that the usual space environment, for which $\varepsilon>0$ and $P>0$, do not generate the cosmological repulsion forces in homogeneous isotropic universe. In order to describe these forces with the Einstein's equations, the existence of negative-pressure media is theoretically assumed. For example, such as: <<dark energy>> ($P=-\varepsilon $); <<quintessence>> ($-\varepsilon<P<-(1/3)\varepsilon$); <<phantom energy>> ($P<-\varepsilon$) (see, e.g.,\cite{13, 14, 15}). From \eqref{d2a} it is clear that such media are sources of cosmological repulsion forces. The shortcoming of these variants of explanation of those forces is  misunderstanding of their physical sense.

In this paper we propose an alternative introduction of the cosmological repulsion forces to the general theory of relativity. It is not based on  the hypothesis of the existence of negative-pressure media. We consider all the components of the space medium to be ordinary that can be described by modern physics. At the same time we suppose that the Einstein equations are not complete. We believe that they only partly account for the effect on the dynamics of the universe of that part of the energy of the space medium, the change of which is purely geometric. In this paper there is given a precise definition of this part of the energy of the space environment. Proposed to calling it <<wave energy>>. A hypothesis is proposed that the wave energy of the space environment is not only the source of the field of attraction, but also of the field of repulsion. Under this hypothesis the solutions were obtained describing the dynamics of the homogeneous isotropic universe. They explain properly the observations for which the effects of cosmological expansion are significant.
	
The paper is written as follows. The generalized Einstein equations describing the cosmological repulsion forces originating from the wave energy of the space environment are given in section \ref{sc:obob_ur_ein}. In section \ref{sc:obob_ur_fridman} these equations are converted to generalized cosmological Friedman equations describing the dynamics of homogeneous isotropic universe. In section \ref{sc:rel_model} a cosmological model of the universe based on these equations (C-model, C~--- Centrifugal) is offered. The examples of using of the C-model to explain some observational data important for cosmology are contained in section \ref{par9}. In section \ref{par10} there are given the results obtained.

\section{Generalized Einstein equations}
\label{sc:obob_ur_ein}
The modern cosmological models are based on the Einstein equations
\begin{equation} 
\label{eq:class_ur_ein}
R^{\mu\nu}-\frac12 R\,g^{\mu\nu}=\frac{8\pi G}{c^4}T^{\mu\nu}+\Lambda g^{\mu\nu},
\end{equation}
see, e.g., \cite{1, 2, 3, 3.1}. Here $R^{\mu\nu}$ --- the Ricci tensor; $R$ --- the invariant obtained by convolution of this tensor; $\Lambda$ --- the cosmological constant; $G$ --- the gravitational constant, $c$ --- the light velocity. It is assumed that the space environment is ideal and for the stress-energy tensor $T^{\mu\nu}$ used an expression (see, e.g., \cite{1, 3}):
\begin{equation}
\label{eq:Tik_virazenie1}
T^{\mu\nu}=(\varepsilon + P)u^\mu u^\nu - P g^{\mu\nu},
\end{equation}
where $u^{\mu}$ --- the four-dimensional velocity.

To describe the dynamics of homogeneous and isotropic universe,  on the assumption of $\Lambda=0$, equations \eqref{eq:class_ur_ein} can be converted to cosmological Friedman equations \eqref{da}, \eqref{d2a}, see, e.g. \cite{1}. From \eqref{d2a} it follows that conventional homogeneous media cannot accelerate the cosmological expansion of the universe. Assuming that the Einstein equations are not complete, let’s question the absolute correctness of this conclusion. This assumption is based on the following qualitative consideration. 

We consider the space environment as an aggregate of particles having a specific wavelength $\lambda$. The wavelengths of various particles are different. At the same time, we believe that the law of their cosmological change is the same. In the comoving three-dimensional frame of reference formula: $\lambda\sim a$ is. It is purely geometric in nature (see, e.g. \cite [ch.3]{1}). Geometrical interpretation of formula $\lambda\sim a$ was proposed long ago by Pauli \cite {16}, Wheeler \cite{17}.

In the process of the cosmological expansion the wavelengths of particles $\lambda$ increase and this reduces their energy estimated by formula:
\begin{equation}
\label{v5}
\varepsilon_\lambda = h\,c / \lambda, 
\end{equation}
which hereafter referred to as <<wave energy>>. In \eqref{v5} $h$ is the Planck constant. Given that the vast majority of the wave energy in the Universe is contained in photons and neutrinos. The influence of the wave energy of the nonrelativistic component on the dynamics of the Universe is negligible. 

We believe that the decrease of the wave energy during the Universe expansion which has a purely geometric nature, leads to the same increase of that part of the particle energy which in the comoving reference frame is connected with  their <<radial>> expansion. The increase of the kinetic energy of the radial expansion of the space environment is interpreted as occurring under the influence of the cosmological repulsion forces. These forces have a purely geometric nature. They are associated with the change, during cosmological expansion of the universe, of the relationship among the wave energy of the particles and their energy coming from the radial expansion of the space environment. In Friedman equations the description of these forces is missing, therefore Einstein equations do not contain it either \eqref{eq:class_ur_ein}.

On the basis of the above we link the effect of the cosmological forces of expansion of the Universe to the wave energy of the environment. The density of this energy of any relativistic component of the space environment is defined in the associated reference frame by the formula:
\begin{equation}
\label{v6}
\varepsilon_w = n \varepsilon_\lambda = h c n /\lambda,
\end{equation}
where $n$ --- the particle density (photons and neutrinos).

In the homogeneous isotropic expanding Universe $n a^3 = const$, $\lambda\sim a$, and thus the wave energy density $\varepsilon_w$ and the scale of the Universe $a$ are interrelated by:
\begin{equation}
\label{v7}
	\varepsilon_w a^4 = const.
\end{equation}

The change of the wave energy density in the process of expansion of the Universe $\varepsilon_w$, that has a purely geometric nature, can be regarded as resulted from the action of the wave pressure forces $P_w$. From the first law of thermodynamics:
\begin{equation}
\label{1nach}
	d(\varepsilon_wV)=-P_wdV,
\end{equation}
considering that $V\sim a^3$ we find that, when holds \eqref{v7}, pressure $P_w$ should be determined by the formula:
\begin{equation}
\label{v8}
	P_w = \frac 13 \varepsilon_w.
\end{equation}
This formula is similar to that for blackbody radiation \cite{1,2,3}.

The wave and thermal energies of the space environment are closely connected. When the space environment is relativistic, then both energies are identical. In general, the higher the temperature of the space environment the greater the wave energy density, the greater its influence on the cosmological expansion of the Universe. In the limiting case of the cold space environment the wave energy is equal to zero and the action of the centrifugal forces should not appear. At the same time in the early Universe, the effect influence of the wave energy on the Universe dynamics is crucial.

We generalize the Einstein equations with intent, as we believe, to correctly describe the influence of the wave energy on the Universe dynamics observing the standard requirements of the general relativity: the form of recording of the generalized equations should remain covariant; the fundamental conservation laws must be observed. In case the contribution of the wave energy of the space environment in its total energy is negligible, the proposed equations should pass into the Einstein equations.

In order to describe the cosmological repulsion forces spreading the Universe we distinguish the part related to the wave energy and wave pressure of the space environment in the stress-energy tensor \eqref{eq:Tik_virazenie1}. To do this we write it as follows:
\begin{equation}
\label{eq:Tik_virazenie2}
T^{\mu\nu}=T^{\mu\nu}_0 + \bar{T}^{\mu\nu},
\end{equation}
where
\begin{equation}
\label{eq:barTik}
\bar T^{\mu\nu}=(\varepsilon_w+P_w)\,u^\mu u^\nu-P_w g^{\mu\nu}.
\end{equation}

Separating wave part $\bar T^{\mu\nu}$ in tensor $T^{\mu\nu}$  we take into account that not only the full stress-energy tensor is covariantly preserved, but its wave part too. We believe that
\begin{equation}
\label{eq:barTzero}
\nabla_\mu\bar T^{\mu\nu}=0
\end{equation}
is correct and has a purely geometric nature. Having determined the wave energy density $\varepsilon_w$ by \eqref{v6}, we calculate wave pressure $P_w$ with equation \eqref{eq:barTzero}. For example, in the case of a homogeneous isotropic Universe  \eqref{v6} and \eqref{eq:barTzero} imply the validity of  formula \eqref{v8}.

Note that the use of formulas \eqref{v6}, \eqref{v8} that allow to define the wave part of the stress-energy tensor is valid when the temperature of the universe is not too high ($T \leq 3\dot 10^9\,K$) and the process of creation of particles and antiparticles does not affect its dynamics.

For an ideal space environment in the homogeneous isotropic universe
\begin{equation}
\label{eq:Tik_virazenie3}
\bar T^{\mu\nu}= P_{w}(4 u^\mu u^\nu - g^{\mu\nu}).
\end{equation}
The trace of tensor $\bar T^{\mu\nu}$ is zero. For the cold  dust-like space environment $\bar T^{\mu\nu}=0$. 

Tensor $T_0^{\mu\nu}$ is not equal to zero due to that the rest mass of a certain portion of the space environment particles is not zero. In normal astrophysical conditions the contribution of the energy associated with the rest mass of the particles to the total space environment energy is crucial. Under these conditions the part of stress-energy tensor $T^{\mu\nu}_0$ is principal and its wave part $\bar T^{\mu\nu}$ is negligible.

The proposed in this article generalization of the Einstein equations is the following. We believe that the wave part of the stress-energy tensor is not only the source of the gravitational field, but at the same time are source of the field cosmological repulsive forces. With this in mind, the generalized Einstein equations written in the form:
\begin{equation}
\label{eq:obob_ur_ein}
R^{\mu\nu} - \frac 12 R\, g^{\mu\nu} = \frac{8 \pi G}{c^4} T^{\mu\nu} - \frac{8\pi \mathbb C}{c^4} \bar T^{\mu\nu} + \Lambda g^{\mu\nu},
\end{equation}
where $\mathbb C$ --- constant of cosmological repulsive forces.

The necessity of introduction of the cosmological repulsion forces originating from the thermal energy of the space environment into GR was discussed before in \cite{18}. In this paper we has clarified the definition of the source of these forces and found their covariant description.

In the limiting case of the vanishing effect of the matter on the metric properties of space the space dynamics is described by the empty space Einstein equations:
\begin{equation}
\label{eq:Tik_virazenie6}
R^{\mu\nu}- \frac 12 \,R\, g^{\mu\nu}= \Lambda\, g^{\mu\nu}.
\end{equation}

This implies that cosmological constant $\Lambda$ is related to the scalar curvature of the empty four-dimensional space-time $R$ by the formula:
\begin{equation}
\label{lambda}
\Lambda = - \frac 14\, R.
\end{equation}
If we use \eqref{eq:Tik_virazenie6} it is easy to show that the scalar curvature $R$ of the empty space-time is a constant.

In this paper we consider that the solution of \eqref{eq:obob_ur_ein} in the limiting case of the vanishingly small effect of the matter on the metric properties of space should pass into one of possible solutions of equation \eqref{eq:Tik_virazenie6} describing the dynamics of the empty three-dimensional space.

It can be shown that for the empty spaces under the assumption of their homogeneity and isotropy, the solutions of equation \eqref{eq:Tik_virazenie6} with $\Lambda\neq 0$, is exponentially divergent. Their interpretation, within the framework of modern physical theories, is difficult. At the same time, with $\Lambda = 0$, for empty homogeneous spaces, there are two physically reasonable and understandable solutions of this equation. The first of them describes a stationary flat three-dimensional space, the distance between any two points of which remains constant. It defines at the infinite the properties of a stationary environment surrounding material point. The second solution describes an evenly expanding open curved space, the curvature radius of which changes at the velocity of light. We believe that the latter is the ultimate state of the space of the infinitely expanding homogeneous Universe. In view of the above we consider the value of the cosmological constant in this work to be equal to zero.

Describing the dynamics of the universe, we use a two-component approximation. We believe that the space environment consists of two homogeneously mixed components: non-relativistic and relativistic.

The non-relativistic component includes all the components of the space environment, both visible (<<baryonic component>>), and yet invisible (<<dark matter>>). This component consists of particles whose rest mass is much larger than their kinetic energy. It is clustered and currently primary by mass/energy in the Universe. The influence of the wave energy of the nonrelativistic component on the dynamics of the Universe is negligible, except for the early Universe when at very high temperatures the particles, whose rest mass differs from zero, can be regarded approximately as a form of radiation (see, e.g., \cite{2}).

In the relativistic component includes all the components of the space environment, as observables (relic radiation) and not observables (relic neutrinos and possibly something else), the equation of state for which is $P=(1/3)\varepsilon$. This component consists of particles of rest mass is zero or much less than their total energy. We assume that the relativistic component is nonclustered, uniformly distributed in space. At present, its contribution to the total energy of the space environment and its influence on the dynamics of the universe is small (see, e.g., \cite{1,2,9}). At the same time in the early universe, this contribution was primary and namely the relativistic component determines the dynamics of the Universe.

The ratio of the particle concentrations of nonrelativistic $n_M$ and relativistic $n_{rad}$ components, except for the earliest stages of evolution of the Universe, remains constant. According to the observational data, $n_{rad}/n_M\sim 10^9$. From now on signs $M$ and $rad$ will be used as is customary (see, e.g. \cite{9}) to denote the quantities describing the nonrelativistic and relativistic components, respectively.

Given that the vast majority of the wave energy in the Universe, except for the earliest stages of its evolution, is contained in the relativistic component of the space environment, we assume that $\bar T^{\mu\nu}=T^{\mu\nu}_{rad}$. With this in mind, the generalized Einstein equations \eqref{eq:obob_ur_ein}, as applied to the description of the dynamics of the Universe, is approximately written as:
\begin{equation}
\label{eq:obob_ur_ein_2k}
R^{\mu\nu} - \frac12 R\,g^{\mu\nu} = \frac{8 \pi G}{c^4} T^{\mu\nu} - \frac{8 \pi \mathbb C}{c^4} T^{\mu\nu}_{rad}.
\end{equation}

In describing the dynamics of the space environment, in which the processes of creation and annihilation of particles are significant, $\bar T^{\mu\nu}\neq T^{\mu\nu}_{rad}$. This means that equations \eqref{eq:obob_ur_ein_2k} are suited for describing the Universe dynamics only at temperatures $T\le 6\cdot 10^9\,K$ (red shifts $z\le 2\cdot 10^9$) when the processes of creation and annihilation of particles have virtually no effect on the dynamics of the Universe \cite{1,2}. At higher temperatures of the space environment these processes are essential and hence there must be considered.

The equations \eqref{eq:obob_ur_ein} describe not only the gravitational fields of attraction, but also cosmological repulsion fields, which are geometric in their nature. The value of constant $\mathbb C$ can be found in the process of understanding and practical use of these equations.

Due to the additional term on the right side of the equation \eqref{eq:obob_ur_ein} the dynamics of the Universe may be fundamentally different from conventional. There may be no singularities in the solutions describing the dynamics of the Universe, its accelerated expansion in the radiation-dominated (RD) era (see \ref{sc:rel_model}) may have place.

\section{Generalized cosmological Friedman equation}
\label{sc:obob_ur_fridman}
Using generalized Einstein equations \eqref{eq:obob_ur_ein_2k} we transform them into cosmological Friedman equations in the usual way. For two-component space environment they can be written as:
\begin{equation}
	\label{eq:fridman_1_1}
		3\left(\frac{\dot{a}^2}{a^2}+\frac{k\,c^2}{a^2}\right) = 8 \pi G (\rho_M + \rho_{rad}) - 8 \pi \mathbb C \rho_{rad},
\end{equation}
\begin{equation}
	\label{eq:fridman_1_2}
		2 \frac{\ddot{a}}{a} + \left(\frac{\dot{a}^2}{a^2}+\frac{k\,c^2}{a^2}\right) = - \frac{8}{3} \pi G \rho_{rad} + \frac{8}{3} \pi \mathbb C \rho_{rad},
\end{equation}
where $a$ --- the curvature radius of the Universe. Upon receipt of these equations it was believed that ${P_{rad}=(1/3)\varepsilon_{rad}}$, $P_M=0$, $\varepsilon_M=\rho_M\,c^2$, $\varepsilon_{rad}=\rho_{rad}\,c^2$. Parameter $k_0$ characterizes the type of the space geometry. It will be shown that constant $k$ can take three values: $k=+1,-1,0$. The value $k=+1$ realizes the case of a space with positive curvature. The value $k=-1$ corresponds to a space with negative curvature. Flat space corresponds to $k=0$. \cite{1,2,3,3.1}.

Bearing in mind the use of equations \eqref{eq:fridman_1_1}, \eqref{eq:fridman_1_2} to describe the expansion of the Universe, we consider that their solutions in the limit of the vanishingly small effect of matter on the metric properties of space must  pass into one of the solutions of \eqref{eq:Tik_virazenie6}. We assume that such solution is the one which describes the  extension of empty space at the velocity of light.
 	
Friedman equations \eqref{eq:fridman_1_1}, \eqref{eq:fridman_1_2} can be written in the dimensionless form:
\begin{equation}
	\label{p1_47}
		\frac{1}{\bar{a}^2} \left( \frac{d \bar{a}}{d \bar{t}}я \right)^2 = -k \frac{\Omega_{curv}}{\bar{a}^2} + \frac{\Omega_M}{\bar{a}^3} + \frac{\Omega_{rad}}{\bar{a}^4} (1-\alpha),
\end{equation}
\begin{equation}
	\label{p1_48}
		\frac{d^2 \bar{a}}{d \bar{t}^2}= - \frac{\Omega_M}{2 \bar{a}^2} - \frac{\Omega_{rad}}{\bar{a}^3} (1-\alpha).
\end{equation}
where $\bar{a}=a/a_0$, $\bar{t}=t\cdot H_0$, $H_0$~--- the Hubble constant. The dimensionless parameter $\alpha=\mathbb C/G$. From now on sign <<0>> is used to denote the quantities that determine the parameters of the modern Universe.

When writing equations \eqref{p1_47}, \eqref{p1_48}, standard notation is used \cite{9}:
\begin{equation}
	\label{p1_49}
	\Omega_M=\frac{\rho_{M\,0}}{\rho_c},\;
	\Omega_{rad}=\frac{\rho_{rad\,0}}{\rho_c},\;
	\Omega_{curv}=\frac{c^2}{H_0^2a_0^2}.
\end{equation}

The Hubble constant is often written in the form of: $H_0=h\cdot 100\,\mbox{km/s\,Mpc}$. Parameters $\Omega_M$ and $\Omega_{rad}$ define in units of $\rho_c$ the current density of nonrelativistic and relativistic components of the space environment, respectively. The critical density $\rho_c$ is defined by:
\begin{equation}
	\label{p1_50}
		\rho_c= 3 H_0^2 / 8 \pi G = 1.88 \cdot 10^{-29}\,h^2\;\mbox{г}/\mbox{см}^3.
\end{equation}

The solutions of equations \eqref{p1_47}, \eqref{p1_48} satisfy the initial conditions:
\begin{equation}
	\label{p1_51}
		\bar{a}(\bar t_0)=1,\;(d \bar{a}/ d \bar{t})(\bar t_0)=1.
\end{equation}
We assume that instant of time $t_0$ corresponds to the present Universe.

\section{The cosmological model of the universe}
\label{sc:rel_model}
The cosmological model of the Universe based on equations \eqref{p1_47} \eqref{p1_48} we call, for brevity sake, C-model. The parameters of the C-model are:
\begin{equation}
\label{parametr1}
\Omega_{curv}, \; \Omega_M, \; \Omega_{rad}, \; k ,\; \alpha \; \mbox{and} \; h.
\end{equation}
Because of \eqref{p1_47}, \eqref{p1_51} they are related by:
\begin{equation}
\label{parametr2}
-k\, \Omega_{curv} + \Omega_M + \Omega_{rad}\,(1-\alpha) =1.
\end{equation}

For the C-model there can be used different solutions which describe the dynamics of the Universe. For a qualitative analysis of these solutions, the equation \eqref{p1_48} can be written as:
\begin{equation}
\label{8.23}
	\frac{d^2 \bar a}{d \bar t^2}=- \frac{dU_C (\bar a)}{d \bar a},
\end{equation}
where
\begin{equation}
\label{8.24}
	U_C (\bar a)=- \frac 12 \frac{\Omega_M}{\bar a} - \frac 12 (1-\alpha) \frac{\Omega_{rad}}{\bar a^2}.
\end{equation}
Equation \eqref{8.23} is similar to the equation describing one-dimensional motion of a particle in potential field \cite{19.2}. The form of solutions $\bar a(\bar t)$ of equation \eqref{8.23} depends on the type of <<potential>> $U_C(\bar a)$, as well as on the value of <<energy>>
\begin{equation}
\label{8.25}
	\bar E=\frac12\left(\frac{d\bar a}{d\bar t}\right)^2 + U_C(\bar a) = -\frac{k}{2}\Omega_{curv}=\frac E{H_0^2a_0^2},
\end{equation}
being the first integral of this equation.

From \eqref{8.25} it can be seen that parameter $k$, which determines the type of the space geometry, is related with <<energy>> $E$ by the formula:
\begin{equation}
\label{eq:f28}
	k=-2\,E/c^2.
\end{equation}
Formula \eqref{8.25} is considered as the law of conservation of the total energy of the space environment per unit mass. For the solutions describing the limitless expansion of the Universe from \eqref{eq:f28} we see that in this case the parameter value is $k=-1$.

\begin{figure}
\includegraphics [width=14cm]{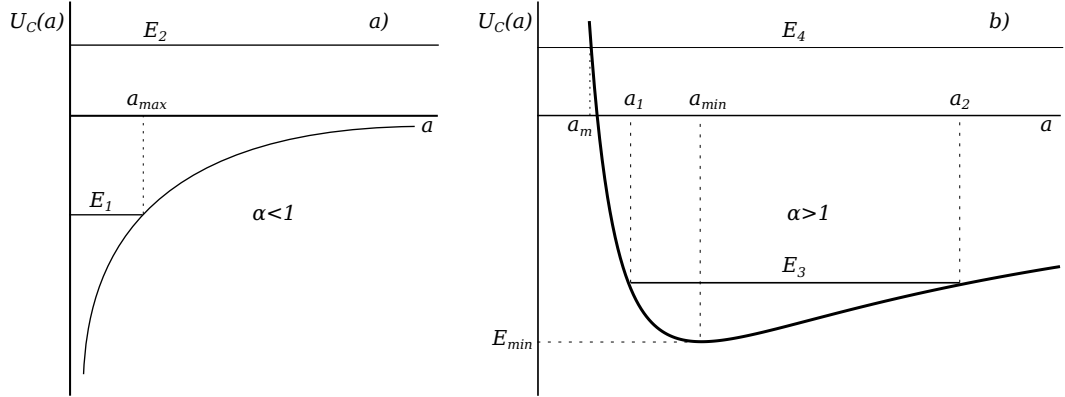} 
\caption{The graphs of function $U_C(a)$, defined by \eqref{8.24}.}
\label{pic3}
\end{figure}

\begin{figure}
\includegraphics [width=14cm]{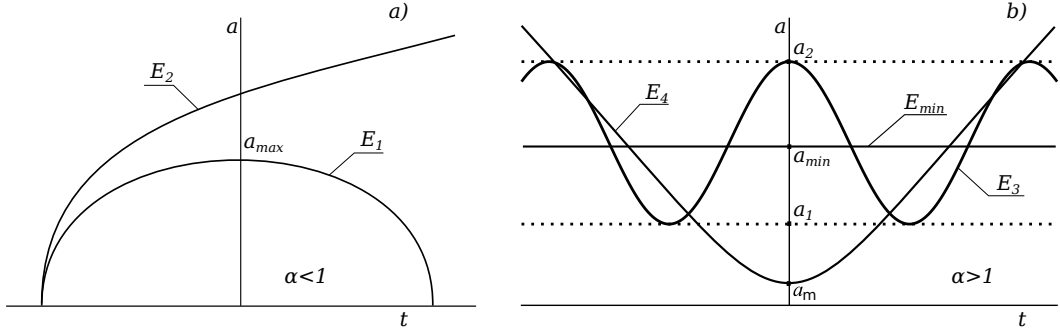} 
\caption{
The qualitative form of the solutions of \eqref{8.23} for different values of parameter $\alpha$ and energy $E$ is shown schematically. The energy levels are shown in Fig. \ref{pic3}.
}
\label{pic4}
\end{figure}

The graphs depicting schematically function $U_C(a)$ are given in Fig. \ref{pic3}. They are qualitatively different for cases $\alpha\leq 1$ and $\alpha>1$. The solutions describing the dynamics of the Universe within the C-model, are schematically depicted in Fig. \ref{pic4}.

When $\alpha\leq 1$ the Universe originates at the instant of <<Big Bang>>. The repulsive forces are always less than the forces of attraction. The expansion takes place with the slowdown. When $E<0$, the Universe is closed and if $E\geq 0$, it is open (see Fig. \ref{pic4} as well.).

In solutions with $\alpha>1$ there is no singularity, there exists a state when the Universe has a minimum scale and maximum temperature. If $\alpha>1$ and $E<0$ the Universe is closed and oscillating. Perhaps, the stable stationary state of the Universe is $E=E_{min}$. The solutions with $\alpha>1$ and $E\geq 0$ describe open Universe (see Fig. \ref{pic4}b.).

Apparently, from the viewpoint of application of the C-model for description of the Universe, those solutions are interesting for which parameter $\alpha$ is greater than unity, but very close to it. According to these solutions, there is no singular state of the Universe, though it may have a very small minimum scale and very large maximum temperature. The Universe expansion is preceded by its contraction.

\section{Explanation of cosmological observations}
\label{par9}
The proposed explanation of the cosmological repulsion forces is only a hypothesis. We suppose that other existing explanations of these forces are also the hypotheses. A substantial argument in support of the correctness of the proposed theory is its ability to explain the cosmological observations.

In this section we shall give an explanation of some of the most important cosmological observations. For comparison to be possible and complete, along with the results obtained in the C-model we give the corresponding results obtained in the $\Lambda$CDM-model. The values calculated in the framework of C- and $\Lambda$CDM- models will be designated by C and $\Lambda$, respectively. In interpretation of the observational data we consider the Universe to be open.

\subsection{The dependence <<apparent magnitude~--- redshift>> for type Ia supernovae}
\label{par9.1}
One way to verify the correctness of cosmological model is based on the comparison of the theoretically calculated in the framework of this model and observable dependences: <<apparent  magnitude ~--- redshift>> for those sources whose absolute luminosity $L$ is assumed to be known, and redshift $z$ to be measured (see, e.g. \cite{4,5,6,7}).

To calculate this dependence, we use the formula:
\begin{equation}
\label{9.11}
	(m-M)(z)=5\lg\left[(1+z)\bar{r}(z)\right]+5\lg\left(c\,H_0^{-1}/l_0\right),
\end{equation}
where $\bar r (z) = r(z) / c H_0^{-1}$, $m=-2.5\lg E+const$, $M=-2.5\lg E_1+const$, $E=L\left/4\pi r^2(z)(1+z)^2\right.$, $E_1=L\left/4\pi l_0^2\right.$, $l_0=10\mbox{пс}$ 
(for details, see, e.g., \cite{1,21}).

The photometric distance $r(z)$ is calculated as follows:
\begin{equation}
\label{rz}
r(z) = a_0 sh \chi (z) , \; \chi (z) = c \int_0^z{\frac{dz'}{a_0 (\dot a /a)_{z'}}}.
\end{equation}
See, e.g., \cite{9, 21}. We assume that the universe is open.

The formula for the photometric distance $\bar r_{\Lambda}(z)$ to the observable object that has a redshift of $z$, can be written as, see, e.g., \cite[ch.4]{9}:
\begin{equation}
\label{89h}
	\bar{r}_\Lambda(z)=\frac{1}{\sqrt{\Omega_{curv}}}\sinh{\int_0^z{\frac{\sqrt{\Omega_{curv}}dz'}{\sqrt{\Omega_{curv}\left(1+z'\right)^2+\Omega_M\left(1+z'\right)^3+\Omega_{rad}\left(1+z'\right)^4+\Omega_\Lambda}}}}.
\end{equation}
Parameters $\Lambda$CDM-model $\Omega_{curv},\; \Omega_M,\; \Omega_{rad}\; \mbox{and}\; \Omega_{\Lambda}$ are related by:
\begin{equation}
\label{90h}
	\Omega_{curv}+\Omega_M+\Omega_{rad}+\Omega_\Lambda=1.
\end{equation}

Usually for interpretation of observations the <<flat $\Lambda CDM$-model>> is used which assume $\Omega_{curv}=0$. The predictions of $\Lambda CDM$-model with $\Omega_{curv}$ significantly differ from zero, contradict to the observations, see, e.g. \cite{9}. The standard mathematical procedure for choosing the theoretically calculated dependence of $(m-M)_\Lambda(z)$, in the best way describes the observational data of supernovae of type Ia, shows that this is valid for $\Omega_M\approx 0.27$, $\Omega_\Lambda\approx 0.73$, see \cite{6, 7, 8}. Taking this into account, we use in the calculations the following values of the parameters of $\Lambda CDM$-model:
\begin{equation}
\label{9.21}
	\Omega_{curv}=0,\;\Omega_M=0.27,\;\Omega_{rad}=\left(4.2/h^2\right)\cdot 10^{-5},\; h=0.7, \;
	\Omega_\Lambda=1-\Omega_M-\Omega_{rad}.
\end{equation}

The values of the parameters of the $\Lambda$ CDM-model, to a great extent, are not the result of direct measurements, but a consequence of their adjustment to correct explanation of various observations.

The formula for the $\bar{r}_C(z)$ in the C-model, taking into account \eqref{p1_47} and \eqref{rz}, can be written as:
\begin{equation}
\label{9.22}
	 \bar{r}_C(z)=\frac{1}{\sqrt{\Omega_{curv}}}\sinh{\int_0^z{\frac{\sqrt{\Omega_{curv}}dz'}{\sqrt{\Omega_{curv}\left(1+z'\right)^2
+\Omega_M \left(1+z'\right)^3
+\Omega_{rad} \left(1-\alpha \right)\left(1+z'\right)^4}}}}.
\end{equation}
The parameters of the C-model are related as follows:
\begin{equation}
\label{9.23}
	\Omega_{curv}+\Omega_M+(1-\alpha)\,\Omega_{rad}=1.
\end{equation}

The evaluating calculations show that the observational cosmological data taken from a wide range of redshifts $z$ ($0<z\lesssim 1000$) are well explained within the framework of the C-model having the following parameters:
\begin{equation}
\label{v39}
	\Omega_M=0.6, \; \Omega_{rad}=(4.2/h^2)10^{-5} , \; \alpha=1, \; h=0.6 , \; k=-1.
\end{equation}

\begin{figure}[h]
\includegraphics [width=14cm]{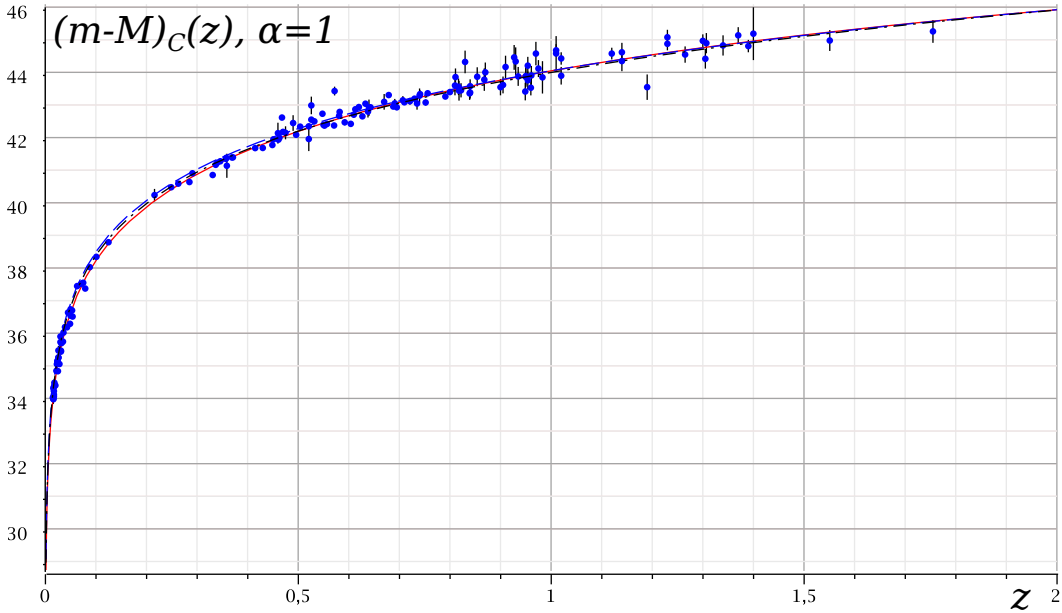} 
\caption{
The relationship $(m-M)_C(z)$ with the values of the parameters \eqref{v39}. For comparison the graphs of dependency $(m-M)_\Lambda(z)$ for <<standard>> values of \eqref{9.21}. The dots in the figure denote the observed values of ${(m-M)(z)}$, the vertical segments designate their measurements errors. The observational data are taken from \cite{6, 7}.
} 
\label{pic5}
\end{figure}

In Fig. \ref{pic5} there is shown a plot of $(m-M)(z)$ calculated in the frameworks of $\Lambda$CDM- and S- models. It is seen that all the graphs properly describe the observed dependence of $(m-M)(z)$ for supernovae of type Ia. Application of these models for explanation of this dependence does not reveal the benefits of one of them.

\subsection{The anisotropy of the relict radiation}
\label{par9.2}
The study of the fine structure of the relict radiation shows that against the uniform background there are minor deviations, (see, e.g. \cite{8,9,10}). Slight variations of the background radiation temperature are seen at the level of a few thousandths of a percent. Following \cite{22,23}, it is considered that they give evidence of weak inhomogeneities of compressions and rarefactions in the cosmic environment in the epoch of recombination.

The brightest spots in the picture of the relict background  are of particular interest.  The observations clearly show an angle of $\Delta\theta$ between the directions in space pointed at the centers of two neighboring bright spots. This angle, with an accuracy of one-two percent, is equal to one degree \cite{8, 9}. The relation between the angular and linear dimensions of the observed object depends on the type of the equations describing the expansion of the Universe, as well as on the parameters of the space environment.

Assuming that the inhomogeneities in the uniform background of relict radiation appeared at the time of recombination, the formula for the angle between the centers of the bright spots can be written as:
\begin{equation}
\label{9.28}
	\Delta\theta=\frac{2\, \bar t_{rec} \left(1+z_{rec}\right)}  {\bar{r} \left(z_{rec} \right)} \cdot \frac{180}{\pi}.
\end{equation}
The angle is taken in degrees (see, e.q. in \cite{21}). In formula \eqref{9.28} $t_{rec}$, $z_{rec}$ --- the age of the Universe and the redshift at the time of recombination, respectively.

The age of the Universe at recombination $\bar t_{rec}$ is found from the condition:
\begin{equation}
\label{9.29}
	\bar{a}\left(\bar t_{rec}\right)=a\left(\bar t_{rec}\right)/a_0=1/\left(1+z_{rec}\right).
\end{equation}
We find function $\bar{a}(\bar{t})$ by solving the equations describing the dynamics of the Universe.

In the $\Lambda CDM$- model, for parameters \eqref{9.21} and $z_{rec}=1000$, we obtain for the recombination time $t_{rec}=4.4\cdot 10^5\,\mbox{years}$ and for the angle $\Delta_\Lambda\theta=1.09^\circ$. The estimated angle value is consistent with the observed one.

For the parameters \eqref{v39} in the C-model, we obtain:
\begin{equation}
	\Omega_{curv}\approx 0.4,\;t_{rec} \approx 4.75 \cdot 10^{5}\,\mbox{years},\;\Delta_C\theta\approx 0.99^\circ.
\end{equation}

This calculated value of angle $\Delta_C\theta$ is consistent with the observed one. At the parameters \eqref{v39} the C-model also explains properly the observed dependence $(m-M)(z)$ for supernovae of type Ia, Fig. \ref{pic5}.

\subsection{Age of the Universe}
\label{sc:par9.3}
To determine the age of the Universe there were found solutions of the equations describing the dynamics of the Universe. In the calculations it was assumed that the time of $\bar t=\bar t_0$ corresponds to the present Universe and at $\bar t=0$ it had the minimum size. Given that $t=\bar t\,H_0^{-1}$, the age of the Universe $t_0$ was estimated as follows:
\begin{equation}
\label{9.35}
	t_0=\bar{t}_0\,H_0^{-1}.
\end{equation}

When performing calculations in the framework of $\Lambda CDM$-model, with parameters \eqref{9.21}, it was obtained: $t_0=13.9\cdot 10^9$ years. With parameters \eqref{v39}the calculations within the C-model gave $t_0=13.0\cdot 10^9$ years.

In conclusion, we will note the following.

The online calculations in the $\Lambda CDM$- and C- models are available on our website \url{www.cosmoway.ru} (see section “Simulation”). Changing any of the parameters of these models, one can see how this affects the dynamics of the Universe.

Application of the proposed tool for explanation of the observations within the C-model shows that the parameters of the Universe are such that it expands for quite a long time almost uniformly (at least in the range of redshifts $z\leq 1000$). Close approximation of the C-model for description of the dynamics of the Universe within redshifts  $(0<z\leq 1000)$ is the uniformly expanding Universe model (S-model \cite{21}). It contains only one free parameter - the Hubble constant. This parameter has clear meaning and is accurately measured.

The above explanations of the observations within the C- and S- models are based on clear and simple physical concepts. From the calculations it does not follow that the three-dimensional space is flat, it does not also follow that the present Universe is expanding with acceleration.

\section{Results}
\label{par10}
\begin{enumerate}
	\item 
An assumption has been made that the cosmological repulsion forces are purely geometric in nature. This assumption is based on the following idea. The cosmological expansion of the Universe is accompanied by an increase in the particle wavelength. In the comoving three-dimensional frame of reference the wavelengths of particles $\lambda$  vary proportionally with the scale of the Universe $a$. Decrease in the energy of particles $\varepsilon_\lambda=h\, c/\lambda\sim a/a$, being purely geometric in nature, is accompanied by simultaneous and equal increase in the particle energy defining the rate of cosmological expansion. The increase in the expansion rate in the comoving frame of reference is interpreted as the result of action of the cosmological repulsion forces. There is no description of these forces in Einstein equations and therefore they are not complete.
	
	\item 
To describe cosmological repulsive forces in the framework of GR we allocate the wave part $\bar T^{\mu\nu}$ of the stress-energy tensor. For ideal space environment this tensor is given by:
\begin{equation}
\label{eq:res_barTmunu}
	\bar T^{\mu\nu}=P_w\left(4\,u^\mu u^\nu-g^{\mu\nu}\right).
\end{equation}	
In the homogeneous isotropic Universe wave pressure $P_w$, wave energy density $\varepsilon_w$ and the scale of the Universe $a$ are related as follows:
\begin{equation}
\label{eq:re_PwEw}
	P_w=\frac13\,\varepsilon_w,\;\;\varepsilon_w a^4=const.
\end{equation}
We believe that tensor $\bar T^{\mu\nu}$ is not only the source of the gravitational field, but also the source of the field of the cosmological repulsion forces. With this in mind, the generalized Einstein equations are written in the form:
\begin{equation}
\label{eq:res_OTOeq}
	R^{\mu\nu}-\frac12\,R\,g^{\mu\nu}=\frac{8\,\pi\,G}{c^4}\,T^{\mu\nu}-\frac{8\,\pi\,\mathbb C}{c^4}\,\bar T^{\mu\nu}+\Lambda\,g^{\mu\nu},
\end{equation}
where $\mathbb C$ --- constant of cosmological repulsive forces.

	\item 
	It is shown that in order for the solution of equation \eqref{eq:res_OTOeq} describing the dynamics of the homogeneous and isotropic open Universe, in the limiting case to pass into a physically reasonable solution describing the dynamics of the homogeneous expanding empty space, the value of the cosmological constant must be assumed to be zero.
	
	\item 
	To describe the homogeneous isotropic Universe the generalized Einstein equations are converted to generalized cosmological Friedman equations. We propose a cosmological model based on these equations (C-model). In the C-model the cosmological repulsion forces are inversely proportional to the cube of the Universe scale. Parameter $\alpha=\mathbb C/G$. An plays a key role in this model. It defines the relation of repulsion and attraction forces in the radiation-dominated epoch.
	
	\item 
For $\alpha\leq 1$, according to the C-model, repulsive forces are always less than the forces of attraction. The Universe originates at the Big Bang moment and then expands with deceleration. Depending on the total energy $E$, the Universe can be closed $(E<0)$ or open $(E\geq 0)$. At $\alpha\leq 1$ the solutions describing the C-model contain a singularity in the behavior of the characteristic scale and the thermodynamic parameters defining the properties of the space environment at $a\to 0$.
	
	\item  
If $\alpha>1$ the singularity in the behavior of the solutions describing the C-model is missing. According to solutions with $\alpha>1$, there was a time when the Universe had minimum size and maximum temperature. According to the C-model, for the maximum temperature to be high enough the value of parameter $\alpha$ should be assumed to exceed one only by a very small quantity. As we believe, among the solutions with such value of this parameter there is the one that describes the Universe. The solution describing the dynamics of the Universe without singularity consists of two symmetrical branches. One of them describes the expansion of the Universe, the other, the contraction. According to this solution, the Universe  originates at infinity and at the end of the evolution regresses there again.
	
	\item 
According to the solutions of the C-model with $\alpha=1+\psi$, where $\psi$ is an infinitesimal quantity, at a certain time the Universe had a very small scale and very high temperature. At that time the repulsive forces were defining. Due to these forces the Universe was rapidly expanding. During extending the repulsive forces decreased inversely proportional to the cubed and the gravity forces, to the squared scale of the Universe. With the passage of time the gravity forces became decisive in the expanding Universe. The observational data on the age of the Universe, the CMB anisotropy and the <<apparent magnitude--redshift>> relationship for supernovae of the Ia type are well explained by the solutions of the C-model according to which the Universe, except for a relatively short initial period, is near uniform expansion. The uniformly expanding Universe model \cite{21} is good approximation for description of its dynamics. The cosmological Friedman equations describing this model are extremely simple:
\begin{equation}
\label{zakl1.1}
	\dot a = \gamma\, c,\;\; \ddot a=0,
\end{equation}
where $\gamma$ --- constant.

The S-model has free parameter - the constant $\gamma$ and the Hubble constant $h$. Selecting values for $\gamma$ and $h$, one can clearly explain the known cosmological observations over a wide range of redshifts $z$ ($0<z\lesssim 1300$).
\begin{itemize}
	\item 
	An explanation for the age of the Universe is offered.
	\item 
	An interpretation of the observed <<apparent magnitude--redshift>> relationship for supernovae of Ia type is set out.
	\item 
	An explanation of the observed angular distance between the centres of neighbouring bright spots on the uniform CMB.
\end{itemize}
\end{enumerate}
	
\bigskip

\begin{thanks}
~We express appreciation to Mssrs. Zhilkin~A.G., and Miller~M.L. for helpful discussions. We are grateful to Mr. Shuhman~I.G. for numerous friendly and efficient discussions. 
\end{thanks}


\end{document}